\documentclass[12pt]{article}

\usepackage[margin=1in]{geometry}
\usepackage{amsmath}
\usepackage{graphicx}
\usepackage{booktabs}
\usepackage{longtable}
\usepackage{array}
\usepackage{tabularx}
\usepackage{hyperref}
\usepackage{url}
\usepackage[numbers,super]{natbib}
\usepackage{setspace}
\usepackage{caption}
\usepackage{subcaption}
\usepackage{textcomp}
\usepackage{authblk}

\title{Measurements Automatically Extracted from Zero Echo Time MRI Using Deep
Learning Image Segmentation and Geometric Modeling Agree with Expert
Manual Readings}

\author[1]{Jack Consolini\thanks{Address correspondence to: Jack Consolini, M.S., Department of Radiology and Imaging, Hospital for Special Surgery, New York, NY 10021, USA. Email: consolinij@hss.edu.}}
\author[1]{Eric A. Bogner}
\author[1]{Meghan Sahr}
\author[1]{Matthew F. Koff}
\author[1]{Kevin M. Koch}
\author[1]{Hollis G. Potter}
\affil[1]{Department of Radiology and Imaging, Hospital for Special Surgery, New York, NY, USA}

\date{}

\begin{document}

\maketitle

\newpage
\begin{abstract}
\noindent\textbf{Background:} Computed tomography (CT) remains the
reference for three-dimensional osseous morphometry for patients with
femoroacetabular impingement (FAI) but entails ionizing radiation and
manual measurement. Zero echo time (ZTE) magnetic resonance imaging
(MRI) permits visualization of cortical bone, and FAI angles from ZTE
MRI agree with CT; however, automated angle extraction remains limited.

\medskip
\noindent\textbf{Purpose:} To develop and validate automated FAI angle
computation from ZTE MRI that will have good to excellent agreement to
manual measurements from expert musculoskeletal radiologists.

\medskip
\noindent\textbf{Study Design:} Cross-Sectional study; Level of evidence, 3.

\medskip
\noindent\textbf{Methods:} Pelvic ZTE MRI was acquired on 73 participants
(mean~$\pm$~standard deviation age $36.8 \pm 18.5$~years; 51~women/22~men)
yielding 135~hips for inclusion. An nnU-Net model was trained on 100
manually curated hips to automatically segment the femur, pelvis, and
three osseous landmarks. Custom-developed automated geometric analysis
algorithms computed alpha, femoral neck-shaft, T\"{o}nnis, coronal and
sagittal center-edge, and acetabular version angles from inferred
segmentations. Measurements on 35 test hips were compared with the mean
of manual measurements made by two musculoskeletal radiologists using
intraclass correlation (ICC) and Bland--Altman analysis.

\medskip
\noindent\textbf{Results:} Dice coefficient exceeded 0.96 for bone
segmentation. Landmark segmentation Dice accuracy ranged from
0.65--0.83. Median landmark error was 0.38~mm (femoral head), 0.82~mm
(lateral acetabulum), and $<$2.5~mm (medial acetabulum and greater
trochanter). ICC was excellent for acetabular versions, coronal
center-edge, and T\"{o}nnis ($\text{ICC} \ge 0.82$) but poor for alpha and
femoral neck-shaft. Model versus rater-mean agreement was excellent for
acetabular version, coronal center-edge, and T\"{o}nnis (0.92--0.96),
good for mid-acetabular sagittal center-edge (0.74), and fair for alpha
(0.45) and femoral neck-shaft (0.55). Bland--Altman limits of agreement
for the model were narrower than interrater limits for most angles.

\medskip
\noindent\textbf{Conclusion:} Fully automated quantitative morphometric
assessment from ZTE MRI is feasible and performs comparably to an expert
reader for most coverage and version angles.

\medskip
\noindent\textbf{Clinical Relevance:} This approach may reduce adjunct CT
for preoperative morphometric assessment of FAI and hip dysplasia in
athletes, young active patients, and women of reproductive age,
providing standardized, automated angle measurements from a single
radiation-free MRI examination.

\medskip
\noindent\textbf{Key Terms:} femoroacetabular impingement; hip dysplasia;
magnetic resonance imaging; automated morphometry; hip preservation;
deep-learning
\end{abstract}

\section*{Abbreviations}

\begin{tabular}{ll}
AIR     & Adaptive image receive \\
CT      & Computed tomography \\
DICOM   & Digital imaging and communications in medicine \\
FAI     & Femoroacetabular impingement \\
HIPAA   & Health insurance portability and accountability act \\
ICC     & Intraclass correlation coefficient \\
MRI     & Magnetic resonance imaging \\
NIfTI   & Neuroimaging informatics technology initiative \\
nnU-Net & No-new convolutional network for biomedical image segmentation \\
PACS    & Picture archiving and communication system \\
RANSAC  & Random Sample Consensus \\
SD$_{\text{diff}}$ & Standard deviation of differences \\
YOE     & Years of experience \\
ZTE     & Zero echo time \\
\end{tabular}

\newpage
\section{Introduction}

Femoroacetabular impingement (FAI) and dysplasia of the hip disrupt
normal joint biomechanics and are established contributors to early
osteoarthritis in young and active patients \citep{breighner2019,griffin2016,hale2021,dancy2025,zhang2015,tanzer2004,lorenzon2020,loder2011}.
Abnormal femoral and/or acetabular morphology produces pathologic
contact during hip motion \citep{breighner2019,griffin2016,hale2021}.
Common sub-types include pincer (acetabular overcoverage), cam (femoral
head--neck asphericity), or combined types
\citep{beck2005,li2016pre,philippon2013}.
Hip dysplasia is characterized by insufficient acetabular coverage of
the femoral head, causing edge loading of the shallow socket
\citep{lorenzon2020,loder2011} and is often accompanied by a cam
deformity, compounding symptoms \citep{ferrell2026,heimer2022}.
Abnormal acetabular shape is common in young populations, with
radiographic evidence of FAI-related morphology in 75--95\% of athletes
\citep{philippon2013,siebenrock2011,yepez2017} and 38\% of hips
evaluated for pain \citep{hale2021}. Progressive impingement contributes
to labral and chondral injury and is implicated in early osteoarthritis
\citep{guevara2006,henak2014}, making accurate morphologic
characterization essential to hip-preservation surgical decision-making
\citep{breighner2019,tanzer2004,beck2005}.

Pre-operative assessment relies on radiography for screening and CT for
detailed three-dimensional osseous morphometry, while MRI remains
essential for soft-tissue evaluation of the labrum and articular
cartilage \citep{breighner2019,li2016pre,li2016post,gold2012,goronzy2019}. Angular
measurements, including alpha, femoral neck-shaft, acetabular version,
and center-edge angles, guide operative planning but are limited on
radiographs by projection error and patient positioning
\citep{smith2018,falgout2023,sakai2009,tannast2007}. CT is therefore
widely used for quantitative morphometry and automated angle extraction
has been developed for CT \citep{tayyebinezhad2025}. Automation of CT
morphometry avoids the excess time-consumption and observer bias of
manual measurements \citep{sangeux2015,nouh2008,notzli2002}, however,
adjunct CT delivers ionizing radiation to a population of primarily
young athletes and women of reproductive age
\citep{breighner2019,tanzer2004}.

Zero echo time (ZTE) MRI provides cortical bone contrast analogous to
CT without ionizing radiation \citep{goronzy2019,koretsky2026}, and
manual ZTE morphometry agrees with CT \citep{breighner2019}. ZTE MRI
could provide comprehensive single-exam morphometric and soft-tissue
evaluation, however, fully automated angle extraction has not been
established for ZTE MRI. Prior automated MRI approaches rely on
statistical shape modeling \citep{xia2015,bugeja2022,ewertowski2022},
which may degrade with atypical anatomy. Therefore, the purposes of
this study were to (1) develop a fully automated pipeline that segments
the femur, pelvis, and osseous landmarks from pelvic ZTE MRI and
computes alpha, femoral neck-shaft, T\"{o}nnis, coronal and sagittal
center-edge, and acetabular version angles using geometric modeling; and
(2) evaluate agreement of automated angles with expert manual
measurements. It was hypothesized that automated angles would agree with
independent manual measurements from two experienced musculoskeletal
radiologists at a level comparable to interrater reliability.

\section{Methods}

\subsection{Enrollment}
This prospective and retrospective single-institution cross-sectional
diagnostic accuracy study (Level of Evidence~III) was approved by the
local Institutional Review Board (IRB\# 2015-441 and 2025-1851) and
conducted in compliance with Health Insurance Portability and
Accountability Act (HIPAA). Prospective participants enrolled with
written informed consent. Retrospective participants were identified
from clinically acquired pelvic MRI under a waiver of informed consent
and HIPAA authorization. Inclusion required pelvic ZTE MRI with
adequate image quality for manual segmentation; no participants were
excluded after enrollment. A subset overlap with a prior publication
evaluating manual ZTE--CT agreement \citep{breighner2019}, which did
not include automated angle computation.

Sample size necessary for a held-out test set was estimated
prospectively using Fisher's $z$-transformation \citep{cohen2013} for a
one-sided test ($H_0$: ICC~$= 0$; $\alpha = 0.05$), with Breighner et
al.\ \citep{breighner2019} ZTE interrater ICC values as anticipated
effect sizes. A minimum of 23 hips was required for 95\% power for all
eight angles.

\subsection{Image Acquisition}
\begin{figure}[htbp]
  \centering
  \includegraphics[width=\textwidth]{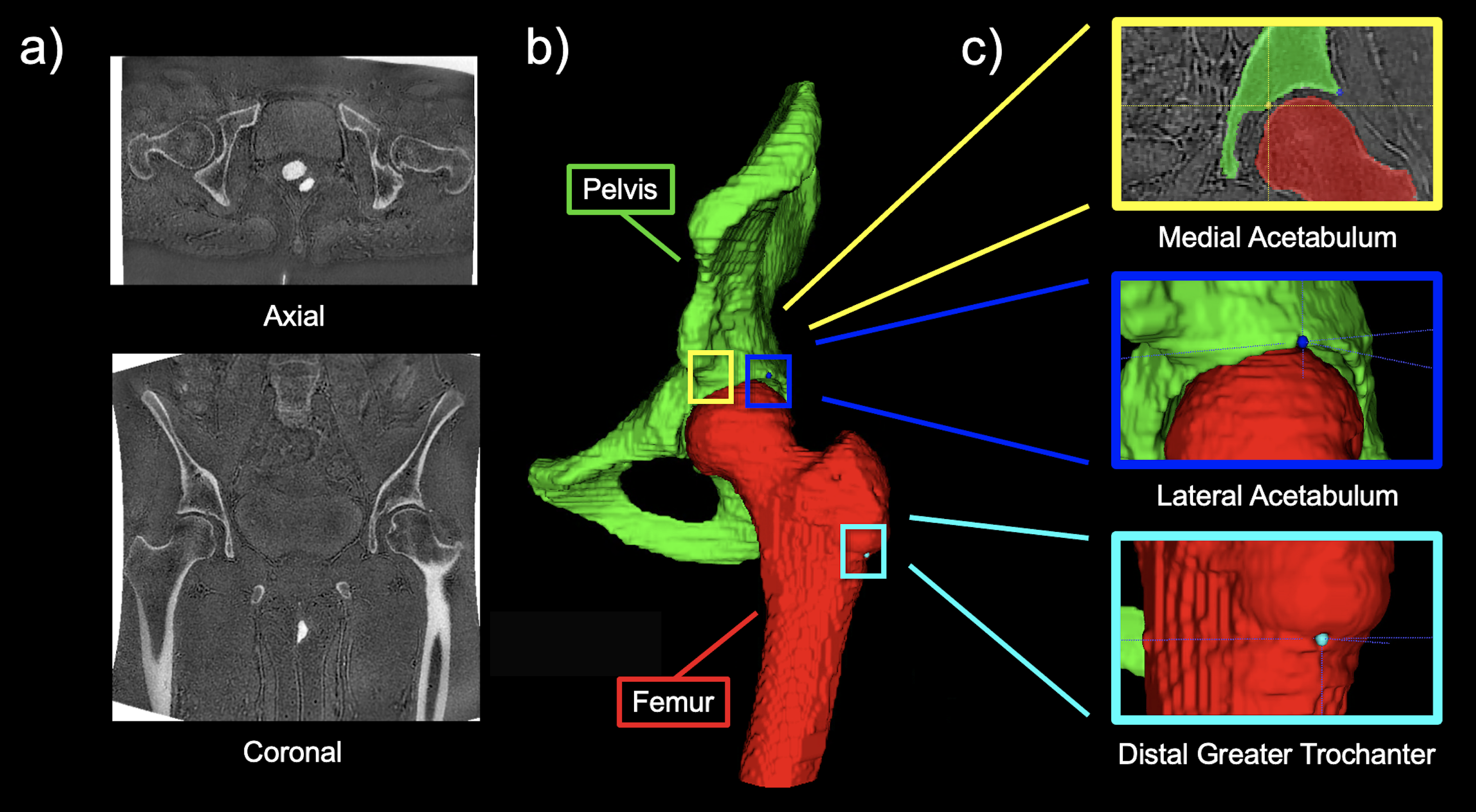}
  \caption{Representative zero echo time (ZTE) hip magnetic
  resonance images (MRI) oZTEo acquisitions and manual segmentation of
  pelvis and femur. (a)~Clinical oZTEo sequence acquired in axial (top)
  or coronal (bottom) orientation without manual post-processing.
  Displayed images received windowing to mimic CT contrast: level
  1026~/~window 1080. (b)~Manual segmentation of femur (red) and pelvis
  (green). (c)~Three fiducial landmarks were placed: lateral acetabulum
  (blue), medial weight-bearing acetabulum (yellow), and distal greater
  trochanter (cyan). The model was trained to predict femoral and pelvic
  masks plus these landmarks.}
  \label{fig:figure1}
\end{figure}

Pelvic ZTE MRI was acquired on a clinical 3.0~T scanner
(SIGNA\textsuperscript{TM} Premier, GE Healthcare, Waukesha, WI) with
either a combination of a 30-channel adaptive image receive (AIR)
anterior array with a 60-channel embedded AIR posterior array coil or
32-channel body array coil for larger patients, or a 32-channel cardiac
coil for smaller patients (GE Healthcare, Waukesha, WI). Patients were
positioned feet-first supine with the feet oriented neutrally and
secured. ZTE was acquired following routine clinical pelvis MRI.
Participants enrolled between 2017 and 2019 received prototype ZTE
sequences post-processed as described by Breighner et al.\
\citep{breighner2019}; subsequent participants received the
FDA-approved oZTEo sequence (GE Healthcare, Waukesha, WI)
(Figure~\ref{fig:figure1}a) with proprietary shading correction. Images
were acquired in the axial or coronal plane.

The acquisition parameters for the prototype ZTE sequences were as
follows: echo time, 0~ms; repetition time, 425--528~ms; flip angle,
1\textdegree; receiver bandwidth, $\pm$62.5~kHz; number of excitations,
4; field of view, 36--44~cm; slice thickness, 1.1--1.5~mm; number of
slices, 100--200; acquisition matrix, $320 \times 320$; and scan time,
$\sim$5~minutes. The acquisition parameters for the oZTEo sequence were
as follows: echo time, 0.016~ms; repetition time, 504.73~ms; flip
angle, 1\textdegree; receiver bandwidth, $\pm$83.33~kHz; field of view,
32--38~cm; slice thickness, 1.0--1.3~mm; number of slices, 180--320;
acquisition matrix, $320 \times 320$; and scan time, $\sim$5--7~minutes.

\subsection{Manual Angular Measurements}
Standard measurements of hip morphology were derived for held-out test
hips by board-certified musculoskeletal radiologists with over 20 and
10 years of experience (YOE), radiologist~1 (E.A.B.) and
radiologist~2 (M.S.), respectively. Angles measured were alpha, femoral
neck-shaft, coronal and mid-acetabular sagittal center-edge, T\"{o}nnis,
and acetabular version (at 1, 2, and 3 o'clock positions) angles. These
measurements constitute the local clinical standard for assessment of
FAI morphology with CT \citep{breighner2019}. Images were provided
within the radiologists' native picture archiving and communication
system (PACS) viewer, Sectra IDS7 (Sectra Medical Systems,
Link\"{o}ping, Sweden). All angles were measured to the nearest 0.1\textdegree.
Manual measurements served as the reference standard for clinical
validation of the automated pipeline.

\subsection{Automated Segmentation Model Development}

\subsubsection{Manual Segmentation of Femur and Pelvis}
The entire bone (trabecular and cortical) of the femur and pelvis was
manually segmented in ITK-SNAP (v3.8.1) \citep{yushkevich2006} from
all hips by a biomedical engineer with greater than 3~YOE (J.C.) and
reviewed as needed by a biomedical engineer with greater than 18~YOE
(M.F.K.) and a board-certified radiologist with greater than 30~YOE
(H.G.P.) (Figure~\ref{fig:figure1}b). High bone-to-soft-tissue contrast
on ZTE images facilitated delineation. Bilateral exams were split at
the midline so each hip could be assigned independently to training,
validation, or test sets and to support acetabular version measurement.

\subsubsection{Manual Identification of Fiducial Landmarks}
Three fiducial landmarks were placed as separate labels within the bone
masks: lateral acetabulum, medial weight-bearing acetabulum, and distal
greater trochanter (J.C.; reviewed by E.A.B.)
(Figure~\ref{fig:figure1}c). Landmarks were dilated to uniform 5-mm
spheres.

\subsubsection{Auto-segmentation Model Training}

\medskip
\noindent\textit{Data Preparation}: Segmentation was performed using no-new convolutional network for
biomedical image segmentation (nnU-Net) \citep{isensee2021}. Digital
imaging and communications in medicine (DICOM) volumes were converted to
neuroimaging informatics technology initiative (NIfTI) format,
standardized to left-posterior-inferior orientation, and cropped to the
side of interest. Six labels were defined: background (0), femur (1),
pelvis (2), lateral acetabulum (3), medial weight-bearing acetabulum
(4), and distal greater trochanter (5).

\medskip
\noindent\textit{Training}:
The default nnU-Net 3D full-resolution configuration was used. Training
employed a weighted composite of cross-entropy and Dice loss with class
weights of 3 for femur and 15 for fiducial labels relative to pelvis.
Five-fold cross validation with an 80/20 train/validation split (per
fold) was employed. Training continued until femur and pelvis mean
validation Dice exceeded 0.90 and fiducial Dice exceeded 0.75 (or
overall Dice exceeded 0.80), or 1000 epochs. Training and inference
were performed on a single NVIDIA Tesla T4 GPU (16~GB VRAM) on an AWS
Elastic-Compute Cloud g4dn.4xlarge instance running Ubuntu 22.04 with
4~CPUs across 2~cores, 125~GB of RAM, and CUDA-enabled (v12.8).

\medskip
\noindent\textit{Inference}:
Final segmentations were generated by using an ensemble of all
five-fold-specific models using sliding-window prediction with a tile
step size of 0.25 (nnU-Net default, 0.5), Gaussian fusion, and
mirror-based test-time augmentation. If fiducial labels were absent
from the ensemble mask, fold-specific predictions were evaluated for
anatomically plausible placement (e.g., distal greater trochanter in
contact with femur) before imputation into the final mask. Predictions
received automated post-processing island removal for femur and pelvic
labels, with only the largest connected regions retained.

\subsection{Automated Angle Computation}
All angles were computed in world coordinates (millimeters) from
predicted segmentations. All described calculations assume standard
Left-Posterior-Inferior-oriented NIfTI data converted to NumPy arrays
via \texttt{sitk.GetArrayFromImage()} \citep{lowekamp2013}. The axial
(Superior--Inferior) axis, coronal (Anterior--Posterior) axis, and
sagittal (Right--Left) axis are identified at runtime from the image
direction cosine matrix by finding the column most aligned with the
corresponding world directions.

\subsubsection{Landmark and Center-of-Mass Localization}
The center of mass of each predicted label was calculated in voxel
space using\\
\texttt{scipy.ndimage.center\_of\_mass} \citep{virtanen2020}
and converted to world coordinates using the image affine defined by
origin, spacing, and direction cosines.

\subsubsection{Femoral Reference Points}
Two femoral reference points were derived from the femur mask (label~1)
and distal greater trochanter landmark (label~5). The femur canal
center is identified on the greater trochanter axial slice. The distal
shaft center is identified on the most distal axial slice in the distal
5\% of the volume whose femur voxel count was $\ge$95\% of the mean
count across a 10-slice sliding window. A conditional fallback was in
place requiring the most distal slice contains $>$10 femur voxels.
Slice-wise 2D centers of mass were converted to world coordinates.

\subsubsection{Femoral Head Center}
The femoral head center was estimated by robust Random Sample Consensus
(RANSAC) sphere fitting to the proximal femoral surface. The femur was
divided at the greater trochanter; the proximal half closer in axial
position to the medial acetabulum (label~4) was retained. Surface
points were obtained by binary erosion. Candidate spheres were fit from
random 4-point subsets over 500 iterations with 3-mm inlier tolerance.
The best model was refined by nonlinear least squares on inliers.

\subsubsection{Automated Angle Definitions}
Angles were defined by the angle between vector pairs in world space
using the dot product.

\medskip
\noindent\textit{Femoral neck-shaft angle}: Angle between the neck axis
(greater trochanter to femoral head center) and shaft axis (greater
trochanter to distal femur center).

\medskip
\noindent\textit{T\"{o}nnis angle}: Angle between the inter-acetabular line
and a horizontal reference constructed at the lateral acetabulum
(label~3) coronal/sagittal coordinates and medial acetabulum (label~4)
axial coordinate; signed inferiorly when the lateral acetabulum lay
below the medial acetabulum.

\medskip
\noindent\textit{Coronal center-edge angle}: Angle between the femoral
head center-to-lateral acetabulum vector and a vertical reference at
the lateral acetabulum axial level and femoral head coronal/sagittal
coordinates.

\medskip
\noindent\textit{Mid-acetabular sagittal center-edge angle}: In the
sagittal plane through the femoral head center, an anterior-inferior
pelvic guide point was identified on the femoral-head sagittal slice.
From the head center, 181~rays spanning 0\textdegree\ to 80\textdegree\
were cast in the sagittal plane; the sourcil was the first pelvic
contact on the ray best aligned with the guide direction (largest
vertical angle used to break ties). The angle was measured between the
superior-inferior axis and the head-to-sourcil vector.

\medskip
\noindent\textit{Acetabular version (1, 2, 3)}: Axial slice levels were
placed at clock-face positions 30\textdegree, 60\textdegree, and
90\textdegree\ from the femoral head equator toward the superior pole.
At each level, posterior-lateral and anterior-lateral acetabular cusps
were identified after splitting the pelvis mask at the femoral head
coronal index. Version was measured relative to a reference line
orthogonal to the line connecting ipsilateral and contralateral
posterior pelvic points; retroversion was assigned a negative value.

\medskip
\noindent\textit{Alpha angle}: Radial planes containing the femoral neck
axis were rotated in 2\textdegree\ increments about the neck. On each
plane, anterior head-neck junction points were identified along the
femoral surface mesh; the plane with the lowest junction-band curvature
variation was selected. The control point was the farthest anterior
junction-band point from the head center, with optional distal
extension when a head-cap gate indicated spherical-cap clustering. Alpha
was the angle between the neck axis and the head-center-to-control-point
vector.

\subsection{Statistical Analysis}

\subsubsection{Model Performance}
Held-out test segmentations were evaluated with Dice similarity
coefficient, Jaccard index, 95th-percentile Hausdorff distance, and
relative volume difference versus ground truth. Results are reported as
mean~$\pm$~standard deviation [minimum, maximum] per structure.

\subsubsection{Fiducial Localization}
Landmark center-point error was defined as the 3D Euclidean distance
(mm) between predicted and ground-truth centers for each fiducial
label; femoral head center error used the RANSAC-derived head center.
Error is summarized by landmark.

\subsubsection{Clinical Validation}
Angles derived from manual measurements from two board-certified
musculoskeletal radiologists (Rater~1~=~E.A.B. and Rater~2~=~M.S.)
were compared with automated ensemble angles on test hips. Reliability
was assessed with two-way random-effects, single-measurement,
absolute-agreement ICC for inter-rater (Rater~1 versus Rater~2),
intra-rater (Rater~1 repeat reads), and model versus rater-mean
agreement. ICC was interpreted as: $<$0.40 poor, 0.40--0.59 fair,
0.60--0.74 good, 0.75--1.00 excellent \citep{cicchetti1994}, with
statistical significance set at $\alpha = 0.05$.

Bland--Altman analysis \citep{bland1986} provided comparison of bias
(mean difference) and 95\% limits of agreement, computed as 1.96 times
the standard deviation of differences (SD$_{\text{diff}}$) for
interrater (Rater~1 versus Rater~2) and model versus rater-mean.
Difference was defined as Rater~1~$-$~Rater~2 and
Model~$-$~mean(Rater~1, Rater~2), respectively. Statistical analyses
were performed in Python using SciPy (v1.13.1), NumPy (v1.26.4),
pandas (v2.3.3), and Pingouin (v0.6.0).

\section{Results}

\subsection{Participant Demographics}
In total, 73 participants (51 women, mean~$\pm$~SD [range] age:
$36.8 \pm 17.7$ [14.5--82.9]~years, BMI: $23.2 \pm 4.0$
[17.5--40.3]~kg/m$^{2}$; men: age $38.4 \pm 20.6$ [15.9--76.8]~years,
BMI: $28.2 \pm 5.9$ [19.6--42.5]~kg/m$^{2}$) and 135 hips were
included (Table~\ref{tab:demographics}). One hundred hips were allocated
to model training and cross-validation. Manual angle validation was
performed on 35 held-out test hips by two musculoskeletal radiologists;
one radiologist repeated measurements on 10 test hips for intra-rater
assessment.

\begin{table}[htbp]
\centering
\caption{Participant Demographics}
\label{tab:demographics}
\resizebox{\textwidth}{!}{%
\begin{tabular}{llcc}
\toprule
Metric & Sex & N (\%) & Mean $\pm$ SD [Range] \\
\midrule
Age    &          & 73 & $36.8 \pm 18.5$ [14.5--82.9] \\
       & Female   & 51 (70.0) & $36.2 \pm 17.7$ [14.5--82.9] \\
       & Male     & 22 (30.0) & $38.4 \pm 20.6$ [15.9--76.8] \\
\addlinespace
BMI    &          & 73 & $24.7 \pm 5.2$ [17.5--42.5] \\
       & Female   & 51 (70.0) & $23.2 \pm 4.0$ [17.5--40.3] \\
       & Male     & 22 (30.0) & $28.2 \pm 5.9$ [19.6--42.5] \\
\addlinespace
Ethnicity &       & 73 & \\
       & Hispanic or Latino          & 4 (5.5)   & \\
       & Not Hispanic or Latino      & 66 (90.0) & \\
       & Unknown                     & 3 (4.5)   & \\
\addlinespace
Race   &          & 73 & \\
       & American Indian or Alaska Native & 0 (0.0) & \\
       & Asian                            & 3 (2.7) & \\
       & Black or African American        & 1 (1.4) & \\
       & Native Hawaiian or Other Pacific Islander & 0 (0.0) & \\
       & White                            & 61 (83.6) & \\
       & Other/Unknown                    & 9 (12.3)  & \\
\addlinespace
Enrollment & & 73 & \\
       & Prospective  & 37 (50.7) & \\
       & Retrospective & 36 (49.3) & \\
\addlinespace
Bilateral Exam & & 73 & \\
       & Unilateral & 9 (12.3)  & \\
       & Bilateral  & 64 (87.7) & \\
\addlinespace
ZTE Acquisition & & 73 & \\
       & Prototype & 23 (31.5) & \\
       & oZTEo     & 50 (68.5) & \\
\bottomrule
\end{tabular}}

\smallskip
\footnotesize Note.---N~=~number of participants, BMI~=~body mass index.
Age and BMI are displayed as mean~$\pm$~standard deviation [range]; N
is displayed as count (\%).
\end{table}

\subsection{Segmentation Performance}
On the held-out test set, femur and pelvis segmentation achieved high
overlap (Dice $0.98 \pm 0.01$ and $0.97 \pm 0.01$, respectively)
(Table~\ref{tab:segmentation}). Fiducial labels had lower but acceptable
overlap (lateral acetabulum $0.83 \pm 0.09$; medial acetabulum
$0.65 \pm 0.18$; distal greater trochanter $0.66 \pm 0.22$). Jaccard
index was consistent with Dice, with excellent overlap for bone (femur
$0.96 \pm 0.01$; pelvis $0.94 \pm 0.02$) and moderate overlap for
fiducials ($0.71 \pm 0.12$ lateral acetabulum, $0.51 \pm 0.19$ medial
acetabulum, $0.52 \pm 0.20$ distal greater trochanter). Relative volume
difference was greatest for femur and pelvis ($3.66 \pm 2.13$~cm$^3$
and $4.50 \pm 3.11$~cm$^3$, respectively), whereas fiducial volume
differences remained an order of magnitude smaller (approximately
0.12--0.15~cm$^3$), apart from outlier cases with incomplete fiducial
predictions. Mean femur and fiducial Hausdorff distances remained 1--4~mm,
reflecting small surface offsets on compact landmarks and the more
linear femur geometry. The mean Hausdorff of the pelvis was 12.2~mm,
with greater offset for larger geometries.

\begin{table}[htbp]
\centering
\caption{Auto-segmentation model performance}
\label{tab:segmentation}
{\fontsize{12}{14}\selectfont
\resizebox{\textwidth}{!}{%
\begin{tabular}{clccccl}
\toprule
Label & Structure & $n$ & Dice & Jaccard Index &
  Hausdorff Distance (mm) & Volume Difference (cm$^3$) \\
\midrule
1 & Femur & 35
  & $0.98 \pm 0.01$
  & $0.96 \pm 0.01$ [0.93, 0.98]
  & $3.5 \pm 1.4$ [1.4, 6.7]
  & $3.66 \pm 2.13$ [0.83, 9.13] \\
2 & Pelvis & 35
  & $0.97 \pm 0.01$
  & $0.94 \pm 0.02$ [0.87, 0.97]
  & $12.2 \pm 10.4$ [3.0, 56.6]
  & $4.50 \pm 3.11$ [0.76, 13.25] \\
3 & Lateral Acetabulum & 35
  & $0.83 \pm 0.09$
  & $0.71 \pm 0.12$ [0.40, 0.92]
  & $2.0 \pm 0.8$ [1.0, 5.0]
  & $0.12 \pm 0.99$ [0.01, 0.39] \\
4 & Medial Acetabulum & 35
  & $0.65 \pm 0.18$
  & $0.51 \pm 0.19$ [0.10, 0.84]
  & $3.5 \pm 1.6$ [1.0, 7.8]
  & $0.15 \pm 0.13$ [0.01, 0.63] \\
5 & Distal Greater Trochanter & 35
  & $0.66 \pm 0.22$
  & $0.52 \pm 0.20$ [0.02, 0.84]
  & $3.6 \pm 2.4$ [1.4, 11.2]
  & $0.15 \pm 0.15$ [0.00, 0.73] \\
\bottomrule
\end{tabular}}}

\smallskip
\footnotesize Note.---$n$~=~number of hips, mm~=~millimeters. All metrics are
relative comparisons of ground truth to predicted segmentations before
any prediction clean-up. Cross-validation mean pseudo Dice and test set
Dice score reported as mean~$\pm$~standard deviation. Jaccard index,
95th percentile Hausdorff distance, and relative volume difference
reported as mean~$\pm$~standard deviation [range]. A maximum of 35
hips for each label were possible within the validation and test set,
respectively.
\end{table}

\subsection{Fiducial Localization}
Center-point error was smallest for the femoral head ($0.38 \pm 0.18$~mm;
median 0.38~mm) and lateral acetabulum ($1.03 \pm 0.78$~mm; median
0.82~mm). The medial acetabulum and distal greater trochanter had
greater error ($2.46 \pm 1.29$~mm and $2.54 \pm 1.77$~mm; medians
2.48 and 1.97~mm, respectively), with wider ranges when trochanter
morphology was indistinct.

\subsection{Clinical Validation}

\subsubsection{Interrater Agreement}
\begin{figure}[htbp]
  \centering
  \includegraphics[width=0.85\textwidth,height=0.8\textheight,keepaspectratio]{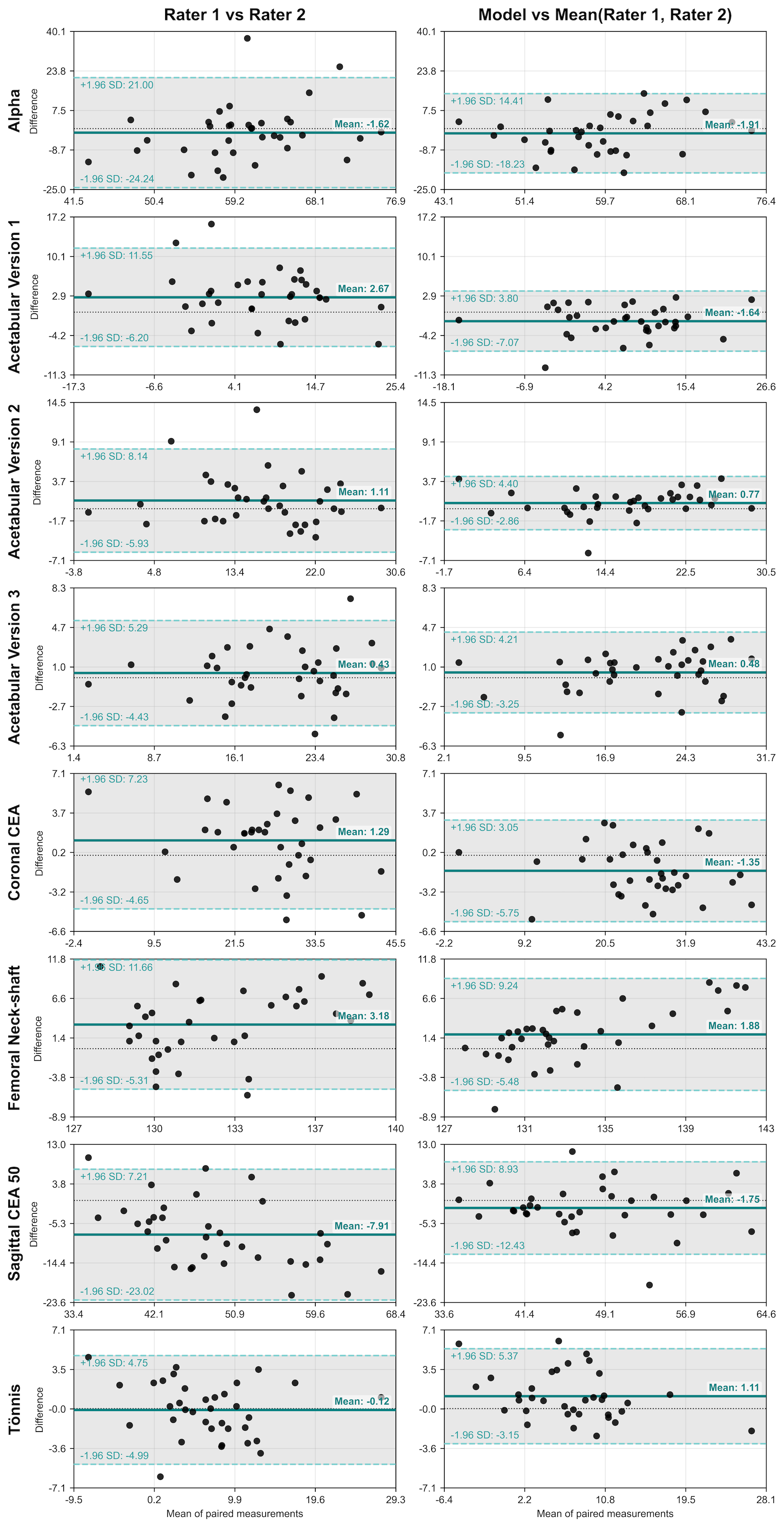}
  \caption{Bland--Altman agreement for automated and manual hip
  angle measurements on zero echo time (ZTE) magnetic resonance images
  (MRI). Rows correspond to individual angular measures; two columns
  compare Rater~1 (E.A.B.) with Rater~2 (M.S.) and the model with the
  mean of Rater~1 and Rater~2. Points show per-hip differences from the
  mean paired value; horizontal lines denote mean bias (teal, solid) and
  $\pm 1.96 \times \text{SD}_{\text{diff}}$ (light teal, dashed), with
  the gray band spanning one $1.96 \times \text{SD}_{\text{diff}}$. The
  zero line is indicated as a thin black dashed line. Tighter point
  clouds indicate closer agreement; alpha and sagittal center-edge at
  mid femoral head height show the greatest dispersion.}
  \label{fig:figure2}
\end{figure}

Acetabular versions 1 (ICC: 0.82), 2 (ICC: 0.86), and 3 (ICC: 0.92),
coronal center-edge (ICC: 0.93), and T\"{o}nnis angle (ICC: 0.93) showed
excellent interrater reliability ($p < 0.001$) (Table~\ref{tab:interrater}
and Figure~\ref{fig:figure2}, Column~1). Mid-acetabular sagittal
center-edge had fair ICC (0.45, $p < 0.001$). Alpha (ICC: 0.20,
$p = 0.129$) and femoral neck-shaft angle (ICC: 0.20, $p = 0.051$) had
poor reliability.

Acetabular versions 1, 2, and 3, coronal center-edge, and T\"{o}nnis
angle had small biases of $+2.67$\textdegree, $+1.11$\textdegree,
$+0.43$\textdegree, $+1.29$\textdegree, and $-0.12$\textdegree,
respectively. Acetabular version 3, coronal center-edge, T\"{o}nnis
angle had smaller $1.96 \times \text{SD}_{\text{diff}}$ of
2.48\textdegree--3.03\textdegree\ than acetabular versions 1 and 2
with $1.96 \times \text{SD}_{\text{diff}}$ of 4.53\textdegree\ and
3.59\textdegree, respectively. Mid-acetabular sagittal center-edge had
large systematic disagreement ($-7.91 \pm 7.71$\textdegree). The widest
limit of agreement was observed for alpha (11.54\textdegree), though
only a small bias $-1.62$\textdegree\ was observed. Femoral neck-shaft
angle had moderate disagreement ($+3.18 \pm 4.33$\textdegree).

\begin{table}[htbp]
\centering
\caption{Interrater agreement assessed via ICC and Bland--Altman}
\label{tab:interrater}
\resizebox{\textwidth}{!}{%
\begin{tabular}{lrcclc}
\toprule
 & & \multicolumn{3}{c}{Intraclass Correlation Coefficient}
 & Bland--Altman (\textdegree) \\
\cmidrule(lr){3-5}
Angle & $n$ & ICC [95\% CI] & $p$ & Reliability
      & Bias $\pm$ 1.96$\times$SD$_{\text{diff}}$ \\
\midrule
Alpha                             & 35 & 0.20 [$-$0.14, 0.49] & 0.129   & Poor      & $-1.62 \pm 11.54$ \\
Acetabular version 1              & 35 & 0.82 [0.58, 0.92]    & $<$0.001 & Excellent & $+2.67 \pm 4.53$  \\
Acetabular version 2              & 35 & 0.86 [0.74, 0.93]    & $<$0.001 & Excellent & $+1.11 \pm 3.59$  \\
Acetabular version 3              & 35 & 0.92 [0.85, 0.96]    & $<$0.001 & Excellent & $+0.43 \pm 2.48$  \\
Coronal center-edge               & 35 & 0.93 [0.85, 0.97]    & $<$0.001 & Excellent & $+1.29 \pm 3.03$  \\
Femoral neck-shaft                & 35 & 0.20 [$-$0.07, 0.47] & 0.051   & Poor      & $+3.18 \pm 4.33$  \\
Mid-acetabular sagittal CE        & 35 & 0.45 [$-$0.04, 0.73] & $<$0.001 & Fair      & $-7.91 \pm 7.71$  \\
T\"{o}nnis                        & 35 & 0.93 [0.86, 0.96]    & $<$0.001 & Excellent & $-0.12 \pm 2.48$  \\
\bottomrule
\end{tabular}}

\smallskip
\footnotesize Note.---ICC~=~intraclass correlation coefficient,
SD~=~standard deviation, CE~=~center-edge. ICC reported as point estimate
[95\% confidence interval]. Bland--Altman: bias~$\pm$~1.96$\times$SD$_{\text{diff}}$
in degrees.
\end{table}

\subsubsection{Intra-rater Agreement}
Acetabular version 3 (ICC: 0.96), coronal center-edge (ICC: 0.96), and
T\"{o}nnis angle (ICC: 0.95) remained excellent on 10 repeated reads by
Rater~1 (all $p < 0.001$), with small biases ($+1.31$\textdegree,
$+1.72$\textdegree, and $-0.54$\textdegree) and
$1.96 \times \text{SD}_{\text{diff}}$ of 1.84\textdegree--2.58\textdegree\
(Table~\ref{tab:intrarater}). Acetabular version 1 showed good
reliability (ICC: 0.83, $p = 0.001$; bias $\pm 1.96 \times
\text{SD}_{\text{diff}}$: $+1.40 \pm 4.05$\textdegree). Acetabular
version 2 (ICC: 0.59, $p = 0.020$; $+3.13 \pm 6.47$\textdegree) and
mid-acetabular sagittal center-edge (ICC: 0.56, $p = 0.013$; $-7.51
\pm 9.61$\textdegree) were fair, though mid-acetabular sagittal
center-edge had a large bias and limits of agreement. Alpha
(ICC: $-0.18$, $p = 0.686$) and femoral neck-shaft angle
(ICC: 0.26, $p = 0.222$) showed poor reliability, with alpha
displaying the widest limits of agreement ($+3.50 \pm 16.70$\textdegree)
and femoral neck-shaft angle moderate spread ($+1.53 \pm 5.19$\textdegree).

\begin{table}[htbp]
\centering
\caption{Intra-rater agreement assessed via ICC and Bland--Altman}
\label{tab:intrarater}
\resizebox{\textwidth}{!}{%
\begin{tabular}{lrcclc}
\toprule
 & & \multicolumn{3}{c}{Intraclass Correlation Coefficient}
 & Bland--Altman (\textdegree) \\
\cmidrule(lr){3-5}
Angle & $n$ & ICC [95\% CI] & $p$ & Reliability
      & Bias $\pm$ 1.96$\times$SD$_{\text{diff}}$ \\
\midrule
Alpha                             & 10 & $-0.18$ [$-$0.78, 0.50] & 0.686   & Poor      & $+3.50 \pm 16.70$ \\
Acetabular version 1              & 10 & 0.83 [0.49, 0.96]        & $<$0.001 & Excellent & $+1.40 \pm 4.05$  \\
Acetabular version 2              & 10 & 0.59 [0.05, 0.88]        & 0.020   & Fair      & $+3.13 \pm 6.47$  \\
Acetabular version 3              & 10 & 0.96 [0.81, 0.99]        & $<$0.001 & Excellent & $+1.31 \pm 1.84$  \\
Coronal center-edge               & 10 & 0.96 [0.78, 0.99]        & $<$0.001 & Excellent & $+1.72 \pm 2.39$  \\
Femoral neck-shaft                & 10 & 0.26 [$-$0.40, 0.75]    & 0.222   & Poor      & $+1.53 \pm 5.19$  \\
Mid-acetabular sagittal CE        & 10 & 0.56 [$-$0.02, 0.87]    & 0.013   & Fair      & $-7.51 \pm 9.61$  \\
T\"{o}nnis                        & 10 & 0.95 [0.83, 0.99]        & $<$0.001 & Excellent & $-0.54 \pm 2.58$  \\
\bottomrule
\end{tabular}}

\smallskip
\footnotesize Note.---ICC~=~intraclass correlation coefficient,
SD~=~standard deviation, CE~=~center-edge. ICC reported as point
estimate [95\% confidence interval]. Bland--Altman: bias~$\pm$~1.96$\times$SD$_{\text{diff}}$
in degrees.
\end{table}

\subsubsection{Model vs.\ Rater-Mean Agreement}
Automatically computed acetabular versions 1--3, coronal center-edge,
and T\"{o}nnis showed excellent agreement with rater mean
(ICC: 0.92--0.96, all $p < 0.001$), with small biases
($-1.64$\textdegree\ to $+1.11$\textdegree) and narrower limits of
agreement than interrater (1.85\textdegree--2.77\textdegree)
(Table~\ref{tab:model_vs_rater} and Figure~\ref{fig:figure2},
Column~2). Automated mid-acetabular sagittal center-edge showed good
agreement (ICC: 0.74, $p < 0.001$; bias $\pm 1.96 \times
\text{SD}_{\text{diff}}$: $-1.75 \pm 5.45$\textdegree). ICC for
automated Alpha (0.45, $p = 0.003$) and femoral neck-shaft (0.55,
$p < 0.001$) were improved from interrater ICC. Femoral neck-shaft had
a smaller bias and tighter limits of agreement ($+1.88 \pm 3.75$\textdegree)
while automated alpha had bias $\pm 1.96 \times \text{SD}_{\text{diff}}$
consistent with the interrater analysis ($-1.91 \pm 8.32$\textdegree).

\begin{table}[htbp]
\centering
\caption{Model vs.\ mean(Rater~1, Rater~2) agreement assessed via ICC and Bland--Altman}
\label{tab:model_vs_rater}
\resizebox{\textwidth}{!}{%
\begin{tabular}{lrcclc}
\toprule
 & & \multicolumn{3}{c}{Intraclass Correlation Coefficient}
 & Bland--Altman (\textdegree) \\
\cmidrule(lr){3-5}
Angle & $n$ & ICC [95\% CI] & $p$ & Reliability
      & Bias $\pm$ 1.96$\times$SD$_{\text{diff}}$ \\
\midrule
Alpha                             & 35 & 0.45 [0.15, 0.68]    & 0.003   & Fair      & $-1.91 \pm 8.32$  \\
Acetabular version 1              & 35 & 0.93 [0.80, 0.97]    & $<$0.001 & Excellent & $-1.64 \pm 2.77$  \\
Acetabular version 2              & 35 & 0.96 [0.91, 0.98]    & $<$0.001 & Excellent & $+0.77 \pm 1.85$  \\
Acetabular version 3              & 35 & 0.95 [0.91, 0.98]    & $<$0.001 & Excellent & $+0.48 \pm 1.90$  \\
Coronal center-edge               & 35 & 0.95 [0.87, 0.98]    & $<$0.001 & Excellent & $-1.35 \pm 2.24$  \\
Femoral neck-shaft                & 35 & 0.55 [0.24, 0.75]    & $<$0.001 & Fair      & $+1.88 \pm 3.75$  \\
Mid-acetabular sagittal CE        & 35 & 0.74 [0.54, 0.86]    & $<$0.001 & Good      & $-1.75 \pm 5.45$  \\
T\"{o}nnis                        & 35 & 0.92 [0.82, 0.96]    & $<$0.001 & Excellent & $+1.11 \pm 2.17$  \\
\bottomrule
\end{tabular}}

\smallskip
\footnotesize Note.---ICC~=~intraclass correlation coefficient,
SD~=~standard deviation, CE~=~center-edge. ICC reported as point
estimate [95\% confidence interval]. Bland--Altman: bias~$\pm$~1.96$\times$SD$_{\text{diff}}$
in degrees.
\end{table}

\section{Discussion}

This study demonstrates that a fully automated ZTE MRI pipeline
combining nnU-Net segmentation, fiducial landmark detection, and
geometric modeling can derive hip morphometric angles highly correlated
with dual-expert manual measurements. This is the first study to compute
alpha, femoral neck-shaft, T\"{o}nnis, coronal and sagittal center-edge,
and multiple acetabular version angles directly from ZTE MRI using deep
learning coupled to explicit landmark-based geometry rather than shape
inference alone. The pipeline improves upon previous MRI-native methods
that rely on statistical shape models
\citep{xia2015,bugeja2022,ewertowski2022} that may degrade in the
setting of marked deformity. The fiducial-first approach embeds
radiologist-intended anatomic priors while retaining full automation,
enabling automated review to function closer to an additional rater
rather than a population-based estimation.

Accurate bone segmentation and submillimeter to low single-digit
millimeter fiducial center-point error support bone and fiducial
landmark-based angle computation. Inter-rater reliability using the
automated landmarks was excellent for acetabular version, T\"{o}nnis, and
coronal coverage measures. Model versus mean grades had improved
agreement over expert manual interrater data for mid-acetabular sagittal
center-edge, indicating this standardized, automated method may produce
more reliable results. Poor alpha and femoral neck-shaft measurements
are consistent with previous reporting
\citep{nouh2008,notzli2002,ewertowski2022,liodakis2012,hermann1997} and
model versus mean rater consensus was within the inter- and intra-rater
agreement range. Evaluating the magnitude of difference for femoral
neck-shaft, poor agreement did not indicate clinically significant
differences. The femoral neck-shaft angle has a normal range of
120\textdegree\ to 135\textdegree, thus a 3\textdegree\ to
5\textdegree\ variation within this range may be inconsequential
\citep{breighner2019,sangeux2015}. The alpha angle was the most
challenging measure because manual identification of the radial
cam-defining plane is inherently variable
\citep{nouh2008,notzli2002,ewertowski2022}. The automated method selects
the flattest junction-band radial section under explicit rules, yielding
a reproducible estimate of maximal cam prominence even when numeric
agreement with a specific reader is limited.

When combined with standard of care pelvis MRI, automated ZTE
morphometry could streamline preoperative evaluation of
hip-preservation candidates by consolidating osseous and soft-tissue
assessment in one study, obviating the need for a CT scan and
time-consuming manual hip angle measurements. Post-operative imaging
could also be performed with automated measurements to improve
reliability of measurements between examinations, which previously may
have been performed by raters with varying experience and techniques.
As a radiation-sparing examination, credible FAI morphometry within MRI
enables a greater ability to safely and accurately track FAI progression
in relatively young athletes across seasons. Areas for further study
include determining if the use of automated angles improves radiologist
and referring physician confidence in measurements, as well as cost and
time savings over the current multi-modality imaging workflow.

This study had several limitations. Data were acquired at a single
institution using a single ZTE acquisition approach while other sites
and MRI vendors may utilize other ZTE acquisition and reconstruction
techniques \citep{grodzki2012}. We anticipate that our methods are
fully translatable if images generated with an alternative sequence
undergo semantic segmentation to define the femur and acetabulum.
Second, the cohort was not demographically or athletically
characterized in detail, and deformity subtype was not systematically
classified. Third, most fiducial training labels were placed by a
trained research engineer under radiologist oversight rather than
annotated directly by a radiologist. Fourth, it may be anticipated that
landmark performance will degrade with severe deformity, artifact, or
atypical anatomy, particularly at the medial acetabulum. Further, alpha
and sagittal center-edge depend on algorithmic choices that may not
mirror manual convention. We anticipate that continued enrollment and
additional model training will be able to account for natural
variability in bone shape. Fifth, the intra-rater assessment was
performed for only 10 hips, although our data is similar to previous
experience \citep{breighner2019}. Additional reads may change agreement
statistics, although minor changes are expected given similar
intrarater performance previously reported \citep{breighner2019}.
Finally, the angles were evaluated without family-wise error control.

In conclusion, this study validated a fully automated ZTE MRI pipeline
that segments bone and fiducial landmarks and computes eight hip angles
commonly used to evaluate patients with FAI and hip dysplasia. Model
versus rater-mean agreement was excellent for acetabular versions 1--3,
coronal center-edge, and T\"{o}nnis (ICC: 0.92--0.96), with Bland--Altman
limits comparable to or narrower than interrater limits for most version
and coverage angles. Alpha and femoral neck-shaft angle require cautious
interpretation given fair model versus mean-rater and poor interrater
reliability. This approach supports efficient, radiation-sparing,
single-modality bone morphometric and soft tissue hip evaluation in
hip-preservation patients.


\section*{Acknowledgements}
The authors would like to thank the staff of the Hospital for Special
Surgery Department of Radiology and Imaging for their assistance with
scanning and acquisition of image data. The authors used Claude
(Anthropic, San Francisco, CA), a large language model, to assist with
editing and formatting of this manuscript; all scientific content, data
analysis, interpretation of results, and conclusions are solely the
work of the authors. The authors take full responsibility for the
accuracy and integrity of the submitted work.

\section*{Funding Statement}
The authors received no financial support for the research, authorship,
and/or publication of this article.

\section*{Ethical Statement}
This study was approved by the local Institutional Review Board
(IRB\# 2015-441 and 2025-1851) and conducted in compliance with the
Health Insurance Portability and Accountability Act (HIPAA). Prospective
participants enrolled with written informed consent. Retrospective
participants were identified from clinically acquired pelvic MRI under a
waiver of informed consent and HIPAA authorization.

\section*{Data Sharing Statement}
Data generated or analyzed during the study are available from the
corresponding author by request.

\section*{Author Contributions}
\textbf{Jack Consolini:} Conceptualization, Data curation, Formal
analysis, Investigator, Methodology, Project administration, Software,
Supervision, Statistical Analysis, Validation, Visualization, Writing
-- original draft, Writing -- review and editing.\\
\textbf{Eric A. Bogner:} Formal analysis, Investigation, Methodology,
Validation, Writing -- review and editing.\\
\textbf{Meghan Sahr:} Formal analysis, Investigation, Methodology,
Validation, Writing -- review and editing.\\
\textbf{Matthew F. Koff:} Conceptualization, Data Curation,
Investigation, Project administration, Resources, Supervision, Writing
-- review and editing.\\
\textbf{Kevin M. Koch:} Conceptualization, Funding acquisition, Project
administration, Resources, Supervision, Visualization, Writing --
review and editing.\\
\textbf{Hollis G. Potter:} Conceptualization, Funding acquisition,
Project administration, Resources, Supervision, Visualization, Writing
-- review and editing.

\bibliographystyle{unsrt}
\bibliography{references}

\end{document}